\documentclass[runningheads]{llncs}
\usepackage{graphicx}
\usepackage{threeparttable}
\begin{document}
%
\title{Steps to Knowledge Graphs Quality Assessment}
%
%
\author{Elwin Huaman\inst{1}\orcidID{0000-0002-2410-4977}}
\authorrunning{E. Huaman}
%
\institute{Semantic Technology Institute (STI) Innsbruck, \\
Department of Computer Science, University of Innsbruck, Austria \\
\email{elwin.huaman@sti2.at}}
\maketitle              
\begin{abstract}
Knowledge Graphs (KGs) have been popularized during the last decade, for instance, they are used widely in the context of the web. In 2012 Google has presented the Google's Knowledge Graph that is used to improve their web search services. The web also hosts different KGs, such as DBpedia and Wikidata, which are used in various applications like personal assistants and question-answering systems. Various web applications rely on KGs to provide concise, complete, accurate, and fresh answer to users. However, what is the quality of those KGs? In which cases should a Knowledge Graph (KG) be used? How might they be evaluated? We reviewed the literature on quality assessment of data, information, linked data, and KGs. We extended the current state-of-the-art frameworks by adding various quality dimensions (QDs) and quality metrics (QMs) that are specific to KGs. Furthermore, we propose a general-purpose, customizable to a domain or task, and practical quality assessment framework for assessing the quality of KGs.
\end{abstract}
\section{Introduction}
\label{sec:introduction}
Large amounts of knowledge are used to power web applications, for instance, in 2012 the Google's Knowledge Graph was presented as means to improve the web search results of Google's search engine. In the same year, Wikidata KG is released as crowd-sourced, non-commercial, and open on the web. Furthermore, many other KGs are available on the web, either created semi-automatically (e.g. DBpedia) or automatically (e.g., NELL). However, an issue that comes with KGs is their quality. Incomplete or inconsistent statements are present in KGs, making them unreliable. Assessing the quality of KGs is a challenge since quality is conceived as \textit{fitness for use}.

Quality may involve several dimensions, such as accessibility, accuracy, appropriate amount, and so on. However, in terms of KGs, few works were proposed. To face this challenge, we look at the literature on quality assessment and identified 20 QDs and several QMs based on the Goal Question Metric approach. 

Furthermore, We propose a general-purpose, customizable to a use case or task, and practical quality assessment framework for KGs. Our approach involves (1) identifying users, use case, and KGs, (2) selection of QDs and QMs by means of weights of importance, (3) performing the assessment of QMs, QDs, and total score for a KG, and (4) exploitation of the results.

This paper is structured as follows. Section~\ref{sec:quality-assessment} provides a look into the literature on quality assessment. In Section~\ref{sec:assessing-kg-quality}, we list QDs that are applicable to KGs. A KG quality assessment framework is provided in Section~\ref{sec:kg-quality-framework}. Finally, we conclude in Section~\ref{sec:conclusion}, by summarizing our findings and future work plans.
\section{Quality Assessment}
\label{sec:quality-assessment}
In this section, we discuss the most relevant work on quality assessment over the past years that implies data, information, linked data, and KGs quality. In order to accurately define and measure a quality dimension (QD), it is not enough to review the quality of data or linked data solely. In fact, KG quality needs an overall picture of quality assessment within the years of their popularization. Defining what quality is, within the context of KGs, will depend greatly on whether QDs have been identified for the data consumers~\cite{WangS96}, the data source~\cite{bizer2007,NaumannR00}, and the use case~\cite{FarberBMR18,FenselSAHKPTUW20,ZaveriRMPLA16}.

\begin{table}
\caption{Comparison of quality dimensions.}
\label{tab:quality-dimensions}
\scriptsize
\centering
\begin{tabular}{|p{1.3cm}|p{3cm}|p{7.5cm}|}
\hline
Ref. & Summary & Dimensions \\ 
\hline
Wang and Strong, 1996~\cite{WangS96} & 15 QDs grouped in 4 categories: Accessibility (2), Contextual (5), Intrinsic (4), Representational (4) & Accessibility, Accuracy, Appropriate amount, Believability, Completeness, Concise representation, Consistent representation, Ease of understanding, Interpretability, Objectivity, Relevancy, Reputation, Security, Timeliness, Value added \\ 
\hline
Naumann and Rolker, 2000~\cite{NaumannR00} & 22 QDs grouped in 3 categories: Object criteria (9), Process criteria (6), Subject criteria (7) & Accuracy, Amount of data, Availability, Believability, Completeness, Concise representation, Consistent representation, Customer support, Documentation, Interpretability, Latency, Objectivity, Price, Relevancy, Reliability, Reputation, Response time, Security, Timeliness, Understandability, Value added, Verifiability \\ 
\hline
Bizer, 2007~\cite{bizer2007} & 16 QDs grouped in 4 categories: Accessibility (2), Contextual (7), Intrinsic (4), Representational (3) & Accessibility, Accuracy, Appropriate amount, Believability, Completeness, Concise representation, Consistency, Consistent representation, Interpretability, Objectivity, Offensiveness, Relevancy, Response time, Timeliness, Understandability, Verifiability \\ 
\hline
Zaveri et al., 2016~\cite{ZaveriRMPLA16} & 18 QDs grouped in 4 categories: Accessibility (5), Contextual (4), Intrinsic (5), Representational (4) & Accuracy, Availability, Completeness, Concise representation, Conciseness, Consistency, Interlinking, Interoperability, Interpretability, Licensing, Performance, Relevancy, Security, Syntactic validity, Timeliness, Trustworthiness, Understandability, Versatility \\ 
\hline
F\"{a}rber et al., 2018~\cite{FarberBMR18} & 11 QDs in 4 categories: Accessibility (3), Contextual (3), Intrinsic (3), Representational (2) & Accessibility, Accuracy, Completeness, Consistency, Interlinking, Interoperability, Licensing, Relevancy, Timeliness, Trustworthiness, Understandability \\ 
\hline
Fensel et al., 2020~\cite{FenselSAHKPTUW20} & 23 QDs & Accessibility, Accuracy, Appropriate amount, Believability, Completeness, Concise representation, Consistent representation, Cost effectiveness, Ease of manipulation, Ease of operation, Ease of understanding, Flexibility, Free of error, Interoperability, Objectivity, Relevancy, Reputation, Security, Timeliness, Traceability, Understandability, Value added, Variety \\
\hline
Hogan et al., 2021~\cite{Hogan21} & 10 QDs in 4 categories: Accuracy (3), Coverage (2), Coherency (2), Succinctness (3) & Accuracy, Coverage, Coherency, Succinctness, Interoperability, Licensing, Relevancy, Timeliness, Trustworthiness, Understandability \\ 
\hline
\end{tabular}
\end{table}
Table~\ref{tab:quality-dimensions} shows an overview of the quality dimension proposed by authors who have a perspective on data~\cite{WangS96}, information~\cite{bizer2007,NaumannR00}, linked data~\cite{FarberBMR18,ZaveriRMPLA16}, and KGs~\cite{FenselSAHKPTUW20} quality. 
The most cited work from this group has been proposed by Wang and Strong~\cite{WangS96}, who empirically collected and organized QDs from data consumer perspectives. They point out the importance of measuring quality to an extent to which data is accessible, contextually dependant on the task at hand, clearly represented, and intrinsically largely correct and complete.

Naumann and Rolker~\cite{NaumannR00} group QDs into 3 categories (subject, predicate, object). The subject (or subject-criteria) implies the user perspective to evaluate specific QDs, the predicate (or process-criteria) implies mainly the process of querying (e.g., availability), and the object (or object-criteria) involves the knowledge source itself (e.g., completeness). Compared with Wang and Strong, Naumann and Rolker highlight the importance of availability, price, customer support, documentation, latency, reliability, response time, understandability, and verifiability QDs. 

Bizer~\cite{bizer2007} argues that quality is a multidimensional concept that involves 3 main aspects, which are the data, its context, and its trustworthiness
. The author proposes consistency and offensiveness as additional QDs compared with ~\cite{NaumannR00,WangS96}. Furthermore, Bizer contemplates offensiveness as a subjective QD that needs to be taken into account in specific contexts (e.g., cultural, religious).

Zaveri et al.~\cite{ZaveriRMPLA16} discuss quality assessment for linked data extensively. They compare 30 papers focused on quality assessment, from where they identified 18 QDs and several assessment tools. Moreover, compared with \cite{bizer2007,NaumannR00,WangS96}, Zaveri et al., introduce additional QDs such as conciseness, interlinking, interoperability, licensing, performance, syntactic validity, trustworthiness, and versatility.

Farber et al.~\cite{FarberBMR18} study the state of the art on quality assessment and implement 11 QDs, which are evaluated through 34 Quality Metrics (QMs) in total. Moreover, the authors test their framework against publicly available KGs (i.e., DBpedia, Freebase, OpenCyc, Wikidata, and YAGO), analyse their results, and provide a list of recommendations on when to use DBpedia, Freebase, OpenCyc, Wikidata, or YAGO KG.

Last but not least, Fensel et al.~\cite{FenselSAHKPTUW20} provide a list of QDs that aim to assess KGs. Compared with~\cite{bizer2007,FarberBMR18,NaumannR00,WangS96,ZaveriRMPLA16}, the authors propose cost-effectiveness, ease of manipulation, ease of operation, flexibility, free of error, traceability, and variety dimensions. Furthermore, Hogan et al.~\cite{Hogan21} provides a list of QDs described in \cite{ZaveriRMPLA16}.

\begin{table}
\caption{Proposed quality dimensions and metrics.}
\label{tab:quality-goals-questions-metrics}
\centering
\begin{threeparttable}
\begin{tabular}{|p{2.7cm}|p{9cm}|}
\hline
Goal & Question, Metrics, and Type (QN/QL) \\ 
\hline
Accessibility & 
Is the KG (or at least part of it) available (QN), provides an SPARQL endpoint (QN), retrievable (QN), supports content negotiation (QN), and contain a license (QN)? \\
\hline
Accuracy & Is the KG reliable and correct, e.g., syntactically (QN) and semantically (QN)? \\ 
\hline
Appropriate amount & Does the KG contain an appropriate amount (QN) of instances for a specific task? \\ 
\hline
Believability &  Does the KG provide provenance information (QN), is trustworthy (QL), and has not unknown nor empty values (QN)?  \\ 
\hline
Completeness & At which degree is the KG complete regarding data (QN), population (QN), and interlinking (QN)? \\ 
\hline
Concise representation & Is the KG concisely represented by avoiding blank nodes (QN) and reification (QN)\\ 
\hline
Consistent representation & Is the KG consistently represented, e.g., disjoint inconsistencies of classes (QN), inconsistent inverse functional property values (QN), schema restrictions (QN)?  \\ 
\hline
Cost-effectiveness &  Does the KG require extra data at any cost (QL)? \\ 
\hline
Ease of manipulation & At which level does the KG provide documentation (QL)?   \\ 
\hline
Ease of operation & Is it possible to update (QN), download (QN), and integrate (QN) the KG?  \\ 
\hline
Ease of understanding & Is the KG represented using self-descriptive URIs (QN) and in various languages (QN)? \\ 
\hline
Free of error & Does the KG provide correct property values (QN)?  \\ 
\hline
Interoperability &  Is the KG interoperable, e.g., openly available (QN) and uses standard vocabularies (QL)?  \\ 
\hline
Objectivity &  Is the KG objective, e.g., unbiased (QL) and declares provenance information (QN)? \\ 
\hline
Relevancy & Is the KG relevant for the task at hand, e.g., at which level the KG provides knowledge for an specific domain or use case (QL)? \\ 
\hline
Reputation & Is the KG well rated? E.g is the KG well positioned in explicit ratings (QL). \\ 
\hline
Security &  Does the KG provide security mechanisms like digital signature (QN) and KG authentication (QL)?  \\
\hline
Timeliness & At which degree is the KG up to date (QN) and fresh (QN)? \\ 
\hline
Traceability & Does the KG provide mean to verify its provenance (QL) and authenticity (QL)? \\ 
\hline
Variety & At which degree the KG integrates various domain sources (QL)? \\
\hline
\end{tabular}
\begin{tablenotes}
    \item [QN: Quantitatively measured metrics.]
    \item [QL: Qualitatively measured metrics.]
\end{tablenotes}
\end{threeparttable}
\end{table}
Most of the authors, with exception to \cite{WangS96,ZaveriRMPLA16}, do not describe the methodology used to define QDs, yet they were created in the context on which the authors focus. In this paper, we consider the dimensions proposed by the authors listed in Table~\ref{tab:quality-dimensions}. Note that flexibility is highly related to ease of manipulation dimension~\cite{FenselSAHKPTUW20}, understandability is also called ease of understanding~\cite{ZaveriRMPLA16}, and value-added is also considered as part of completeness~\cite{FarberBMR18}. In order to define QDs that are relevant to KGs, we look up into KG's definition, which is characterized by its variety to cover various domains~\cite{FarberBMR18,FenselSAHKPTUW20,Paulheim17}. Furthermore, we consider that some dimensions are getting attention over the last years in the literature, such as cost~\cite{FenselSAHKPTUW20,Lenat95,HuamanF21,WangS96} and traceability~\cite{FenselSAHKPTUW20,HuamanKF2020,HuamanTF21} dimensions. We summarize our findings into 20 QDs that are relevant in the KGs context, see Table~\ref{tab:quality-goals-questions-metrics}.

\section{Assessing Knowledge Graph Quality}
\label{sec:assessing-kg-quality}
In order to operationalize the measurement of the above twenty quality dimensions, we employ the Goal Question Metric (GQM) approach. The GQM approach implies defining (i) a goal, (ii) a set of questions to achieve the goal, and (iii) a set of metrics to answer the questions. Although the GQM approach is used to measure software quality, it has been used in the context of data quality~\cite{Redman0093293} and Linked Open Data~\cite{BehkamalKBJ14}. Furthermore, we classify each metric as being quantitatively (QN) or qualitatively (QL) assessed. QN implies that a value score can be calculated for that metric (e.g., manually, semi-automatically, or automatically) and QL requires user judgement to assess the metric (e.g., user perception w.r.t. a metric).

Table~\ref{tab:quality-goals-questions-metrics} displays the 20 QDs along with questions, metrics, and metric's classification as we understand them. Of course, these definitions might not satisfy every use case or domain. Therefore, they can be adapted and improved.
\section{Knowledge Graph Quality Assessment Framework}
\label{sec:kg-quality-framework}
\begin{figure}
\includegraphics[width=\textwidth]{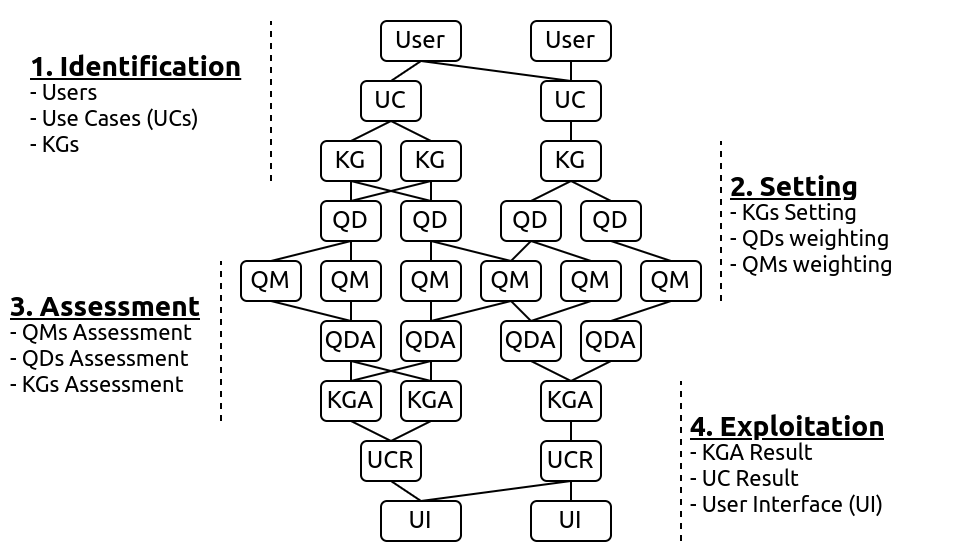}
\caption{Hierarchical decomposition of our KG Quality Assessment Framework.} 
\label{fig:hierarchical-process}
\end{figure}

In order to assess a KG, we propose a user-driven assessment framework that enables users to select quality dimensions and metrics according to their degree of importance (e.g., domain, a task at hand, or use case). Our approach consists of 4 steps: Identification, setting, assessment, and exploitation (see
Fig.~\ref{fig:hierarchical-process}).

\subsection{Identification}
\label{subsec:identification}
In this step, we propose to identify 3 main points: a user, a use case, and a KG.
\begin{enumerate}
    \item[i.] \textit{Users} (namely domain experts) will guide the process of defining the use case, KGs to evaluate, and the weights for QDs and QMs.
    \item[ii.] \textit{Use Cases} (UCs), domain, or tasks at hand need to be understood so that the weights are assigned appropriately. In KGs context, it is important to identify which QDs and QMs are relevant for a domain (see Table~\ref{tab:quality-goals-questions-metrics}).
    \item[iii.] \textit{KGs} must be identified by the users as relevant for the UC, e.g., for Points of Interest (POI) domain, KGs such as DBpedia, Google's Knowledge Graph, and Wikidata may be suggested.
\end{enumerate}



%
\subsection{Setting}
The goal of developing an assessment framework is to obtain a quality status of a KG for a specific UC. For that, the user needs to set up a KG into the framework, then select, based on weights of importance, a set of QDs and QMs to evaluate in the KG. This process involves:
\begin{enumerate}
    \item[i.] \textit{KGs Setting}. It is not a straight forward task, for instance. KGs use different schema for describing a same property, e.g., label\footnote{label is a RDFs property used to describe a name for a subject. \url{https://www.w3.org/2000/01/rdf-schema\#label}}, name\footnote{name is a property used by Schema.org. \url{https://schema.org/name}}, or Q82799\footnote{Q82799 is a property used to describe the name of an instance in the Wikidata KG. \url{https://www.wikidata.org/wiki/Q82799}}. Moreover, there are KGs that do not provide a SPARQL endpoint (e.g. Google's Knowledge Graph), so SPARQL queries must be adapted to a different query format (e.g. API).
    \item[ii.] \textit{QDs Weighting}. Beta-weights (\(  \beta_{i} \)) are defined for selecting QDs, such as \(  \beta_{i} \) defines a weight of importance for each \( d_{i} \) QD and \(  \beta_{i} \in  \left[ 0,1 \right]  \) where \( 0 \) is the minimum degree of importance and a value \( 1 \) is the maximum degree. Furthermore, it must hold:
\[ \sum _{j=1}^{20} \beta _{j}=1  \] 
    \item[iii.] \textit{QMs Weighting}. Alpha-weights (\( \alpha_{i,j} \)) defines a weight of importance for each QM, such as \( \alpha_{i,j_{i}} \) defines a weight of importance for a \( m_{i,j} \) QM and \( \alpha_{i,j} \in  \left[ 0,1 \right]  \) where \( 0 \) is the minimum degree of importance and a value \( 1 \) is the maximum degree. Moreover:
\[ \sum _{j=1}^{j_{i}}\alpha_{i,j}=1  \textrm{ for all } i = 1, ... , 20 \] 
\end{enumerate}
\subsection{Assessment}
In this step, the framework performs 3 main tasks: Assessing the QM, calculating the aggregated score for QDs, and the total aggregated score for a KG. 
\begin{enumerate}
  \item[i.] \textit{QMs Assessment} (QMA) involves the setting and performing of QMs, so they can be performed either manually, semi-automatically, or automatically.
  \item[ii.] \textit{QDs Assessment} (QDA) is represented by $d_{i}(g)$, which calculates the QD aggregated score from the values of each of its QMA for a KG $g$. 
 \[ d_{i}(g) = \sum _{j=1}^{k_{i}}{m_{i,j}}.\alpha_{i,j} \] 
  Furthermore, $k_{i}$ is the number of metrics for the $i^{th}$ QD, $m_{i,j}$ is the score of the $j^{th}$ QM of the $i^{th}$ QD, the alpha-weights ($\alpha_{i,j}$) define the impact of the $j^{th}$ QM score on the $i^{th}$ QD score, and $d_{i}(g)$, $m_{i,j}$, $\alpha_{i,j} \in [0,1]$.
  \item[iii.] \textit{KG Assessment (KGA)} is represented by $T(g)$, which calculates the total aggregated score from the values of each QDA for a KG $g$. 
 \[ T(g) = \sum _{i=1}^{n}{d_{i}(g)}.\beta_{i} \] 
  Furthermore, $n$ is the number of QDs, ${d_{i}(g)}$ is the score of the $i^{th}$ QD of $g$, $\beta_{i}$ represents the weight of the $i^{th}$ QD, and $T(g)$, $d_i(g)$, $\beta_i \in [0,1]$.
\end{enumerate}
As described above, the alpha- and beta- weights select, as well as define, the impact of the QDs and QMs on the overall score of the assessed KG.

\subsection{Exploitation}
As described above, the alpha- and beta- weights affect the overall result of an assessed KG. Therefore, the results can be compared, tuned, and visualized.
\begin{enumerate}
    \item[i.] \textit{KGA result}. The alpha- and beta- weights affect the KGA result. In this step, users can tun the alpha- and beta- weights in order to get desirable results, which might update on the fly.
    \item[ii.] \textit{UC Result} (UCR). The main goal of developing an assessment framework is to obtain suitable KGs for specific use cases. UCR compares the values resulting from the KGAs per UC. For instance, it ranks the results.
    \item[iii.] \textit{User Interface} (UI). It must provide various means by which the calculated results can be accessed, visualized, and easily interpreted. For instance, giving recommendations on when to use such KGs. 
\end{enumerate}
Note that, our framework defines QDs and QMs as we understood them. However, every situation or application might require some adaptation. Also notice that some QM are similar to each other or one QM might be combined in more than one QD, therefore we advise not to select all QD and QM at the same time, e.g., play with the alpha- and beta- weights until you get a desirable result.

%
%
\section{Conclusion and Future Work}
\label{sec:conclusion}
In this paper, we conducted a comprehensive study on quality assessment, which included data, information, linked data, and KGs quality context. We propose 20 QDs that are relevant in the KGs context and several metrics that can be quantitatively or quantitatively measured. Furthermore, we remark the importance of taking into account cost-effectiveness, traceability, and variety QDs. Then, we propose a user-driven assessment framework for evaluating the quality of KGs. Our framework comprises 4 steps: identification, setting, assessment, and exploitation. 
Furthermore, The framework can be used for lighting KGs architects on the basic quality requirements of KGs, and in the development of future KG assessment frameworks.

The next step of our work will be focused on developing the KG assessment framework. Moreover, we will evaluate the performance of the framework and conduct surveys from domain experts and KG researchers to evaluate and improve the proposed QDs, as well as, the framework. Furthermore, the development of a KG assessment framework might be, to some extent, semi-automated, i.e., implementing the metrics that can be quantitatively measured (which also might reduce assessment costs), but it will still be a combination of human-machine efforts as necessary.

On one hand, since there is no unique solution to assess the status of KGs, further research is needed to evaluate the framework in specific use cases. On the other hand, a unified quality assessment approach would improve the applying of quality assessment in KGs and increase the adoption of KGs to worldwide application scenarios.

\textbf{Acknowledgments.} The initial work for this paper was funded by the MindLab project: \url{https://mindlab.ai/}
%

\bibliographystyle{splncs04}
\bibliography{bibliography}

\end{document}